\documentclass[conference]{IEEEtran}
\usepackage{booktabs,makecell}
\usepackage{amsmath,amssymb,amsfonts}
\usepackage{algorithmic}
\usepackage{graphicx}
\usepackage{textcomp}
\usepackage{xcolor}
\usepackage{adjustbox}
\usepackage{gensymb}  
\usepackage{multirow}
\usepackage[utf8]{inputenc}
\usepackage[english]{babel}
\usepackage{gensymb}
\begin{document}
\title{LSTM-based approach to detect cyber attacks on market-based congestion management methods\\}
\author{\IEEEauthorblockN{Omniyah Gul M Khan\IEEEauthorrefmark{1},
Amr Youssef\IEEEauthorrefmark{2},
Ehab El-Saadany\IEEEauthorrefmark{3}\IEEEauthorrefmark{4}, and
Magdy Salama\IEEEauthorrefmark{5}}
\IEEEauthorblockA{\IEEEauthorrefmark{1}Electrical and Computer Engineering, University of Waterloo, Canada, Email: ogulmkha@uwaterloo.ca}
\IEEEauthorblockA{\IEEEauthorrefmark{2}Information Systems Engineering, Concordia University, Canada, Email: youssef@ciise.concordia.ca}
\IEEEauthorblockA{\IEEEauthorrefmark{3}Advanced Power and Energy Center, EECS Department, Khalifa University, Email: ehab.elsadaany@ku.ac.ae}
\IEEEauthorblockA{\IEEEauthorrefmark{4}Electrical and Computer Engineering, University of Waterloo, Email: ehab@uwaterloo.ca}
\IEEEauthorblockA{\IEEEauthorrefmark{5}Electrical and Computer Engineering, University of Waterloo, Email: msalama@uwaterloo.ca}}
\maketitle
\begin{abstract}
Market-based congestion management methods adopt Demand Side Management (DSM) techniques to alleviate congestion in the day-ahead market. Reliance of these methods on the communication layer makes it prone to cyber attacks affecting the security, reliability, and economic operation of the distribution network. In this paper, we focus on Load Altering Attacks that would compromise the operation of market-based congestion management methods. A detection technique is proposed using Long Short-term Memory (LSTM) Recurrent Neural Networks (RNN). IEEE 33 bus system is used as a case study to demonstrate the effectiveness of the proposed technique. An accuracy of 97\% was obtained proving the capability of using LSTM-RNN to detect a load altering cyber attack compromising aggregators in the network.
\end{abstract}

\section{Introduction}
Electrification of the transportation system and space heating, and the increased penetration of Distributed Energy Resources have changed the operating conditions of the distribution system. Distribution System Operators (DSO) are facing congestion problems due to power flow exceeding the network's transfer capability. To manage congestion that could cause voltage violations and/or thermal overloading, possibly damaging devices such as distribution transformers and feeders, market-based congestion management methods have been developed. Such methods utilize existing flexibility in the market through Demand Side Management (DSM), avoiding incurring huge costs to reinforce network assets \cite{article1}. 

Market-based congestion management methods \cite{b1} \cite{o9} employ the communication layer to relay consumer preferences to their respective aggregators, price tariffs from DSO to aggregators, and load schedules from aggregators to the DSO to mitigate congestion. The success of such congestion management methods, hence, relies on the cyber security of the entities involved and their communication links. Ensuring the confidentiality, integrity, and availability of the DSO, aggregators, consumers' meters, and the information exchanged is vital to resolve congestion using DSM.  

Cyber attacks compromising consumers' smart meters and their communication lines have been studied in the literature. \cite{j6} studied the impact of closed-loop load altering attacks compromising the stability of the power system. In \cite{lit1}, three types of attackers were studied, with varying cyber attacking capabilities, that aim to steal electricity using false data injection (FDI) by tampering with the smart meters. On the other hand, in \cite{lit2}, the impact of a cyber attack manipulating the readings of consumers' meters responsible for monitoring renewable DG units to overcharge the utility. 

Attacking aggregators or their communication links with the DSO is another aspect that can be utilized by a cyber attacker to affect the performance of market-based congestion management methods. Aggregators hold a unique position in the distribution network in terms of their connection to their respective DER equipment and the DSO \cite{rev2}. Compromising the cyber security of aggregators could risk the overall security of the grid. Financial losses can be caused as a result of an adversary causing congestions not to be detected in the day-ahead market by the DSO. This would cause unnecessary load shedding that could be avoided if the attack was mitigated. Moreover, cyber attacks can be executed by DSO competitors to fake congestion resulting in higher congestion tariffs to be imposed on the consumers. This would encourage consumers to change their utility provider. Hence, it is critical to study the impact of cyber attacks on aggregators and how to mitigate or detect such attacks. However, no literature exists on studying attacks compromising aggregators explicitly except our work in \cite{omo1}. In \cite{omo1}, the impact of compromising aggregators, as a result of Load Altering Attacks (LAA), on congestion and congestion tariff has been evaluated. But, up to the author's knowledge, no literature exists on detection of such attacks that compromise aggregators load profiles communicated to the DSO affecting congestion in the network. 

Motivated by the research gap, the main contribution of this paper is focused on developing a detection scheme that would determine the presence or absence of a load altering cyber attack compromising aggregators in the distribution network. The detection scheme is executed by the DSO in the day-ahead market on receiving aggregators' load profiles prior to computing the congestion tariff. 

\section{Related Work}
Load Altering Attacks (LAAs) has been previously studied in the literature. In \cite{ami8}, static LAAs compromising Internet connected vulnerable loads were studied. A cost-efficient load protection was then proposed identifying the vulnerable loads that need to be secured without overloading the distribution system. In \cite{j6}, closed-loop dynamic LAA were studied in which non-secured load was cyber-attacked over time in a pre-programmed trajectory based on system frequency. A non-convex pole placement optimization problem is then solved to determine the minimum load required to be secured without impacting power system stability. It should be noted, however, that \cite{ami8}-\cite{j6} were studied based on power system dynamics. 

Detection of FDI attacks utilizing data-based methods has been also studied. Data-based methods' ability to scale to larger systems with low computational cost makes it an attractive technique \cite{det3}. \cite{det1} tested and compared the performance of linear and Gaussian Support Vector Machines (SVM), K-Nearest Neighbours (KNN), and single layer perceptrons as FDI detection models. Similarly, \cite{det2} utilized SVM, KNN, and Extended Nearest Neighbours to detect the most accurate method to detect falsification of measurement and state data. A SVM-based classifier was adopted to detect energy theft attacks as a result of hacking smart meters in \cite{lit4}. 

\section{Case Study System Details}
\begin{figure}[t]
\centering
  \includegraphics[trim={0cm 1cm 0cm 0cm}, width=0.45\textwidth, height=0.2\textwidth]{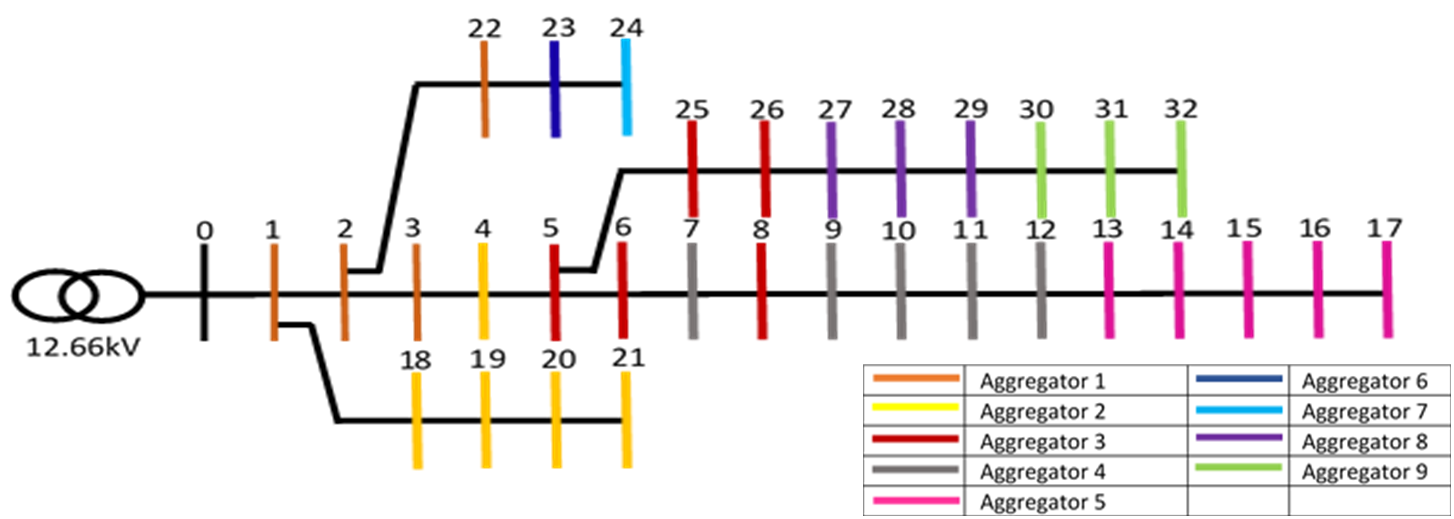}
  \caption{IEEE 33 bus system involving nine aggregators \cite{b24}}
  \label{fig:fig2}
\end{figure}

To simulate load altering attacks on market-based congestion management techniques and the proposed detection model, the IEEE 33 bus system \cite{b24}, illustrated in Figure \ref{fig:fig2}, is adopted. Having 32 load buses, residential consumers are represented using both flexible and base loads. Residential base load profiles are designed to have a maximum load equivalent to the default loading of the IEEE 33 test bus system. On the other hand, residential flexible loads consist of one Heat Pump (HP) and one Electric Vehicle (EV) with a battery size of 36kWh, per residence. Consumers' preferences in terms of their comfortable indoor temperature range and initial State of Charge (SOC) of EV are randomized to be uniformly distributed between $18-24\degree \text{C}$ and $20-30\%$ respectively. The residential charging power of EVs was chosen as 11kW. All the load buses of the IEEE 33 bus system, are assumed to be residential, having both flexible and non-flexible loads, except bus 23 and 24. Having the largest base-load, load bus 23 and 24 were utilized as commercial buses consisting of Fast Charging Direct Current (FCDC) EV charging parking lots of 50kW. Arrival and departure of EVs from the FCDC parking lot were simulated using data obtained from Toronto parking authority \cite{extra}. The absence of an EV from the commercial parking lot was assumed to indicate its presence at its residence. All residential and commercial buses were represented by 9 aggregators having comparable loads, as illustrated in Figure \ref{fig:fig2}. Aggregators 6 and 7 are used to represent commercial buses. Day-ahead electricity prices are obtained from the PJM market via the dataminer tool \cite{pjm}. GAMS \cite{b25} was used for executing the optimizations in this paper. MATLAB was used for determining the congestion tariffs imposed by the DNO to relieve congestion and for implementing the proposed detection scheme.

\section{Impact of LAAs compromising aggregators}
\begin{figure}[b]
\centering
  \includegraphics[trim={0cm 1cm 0cm 0cm}, width=0.4\textwidth]{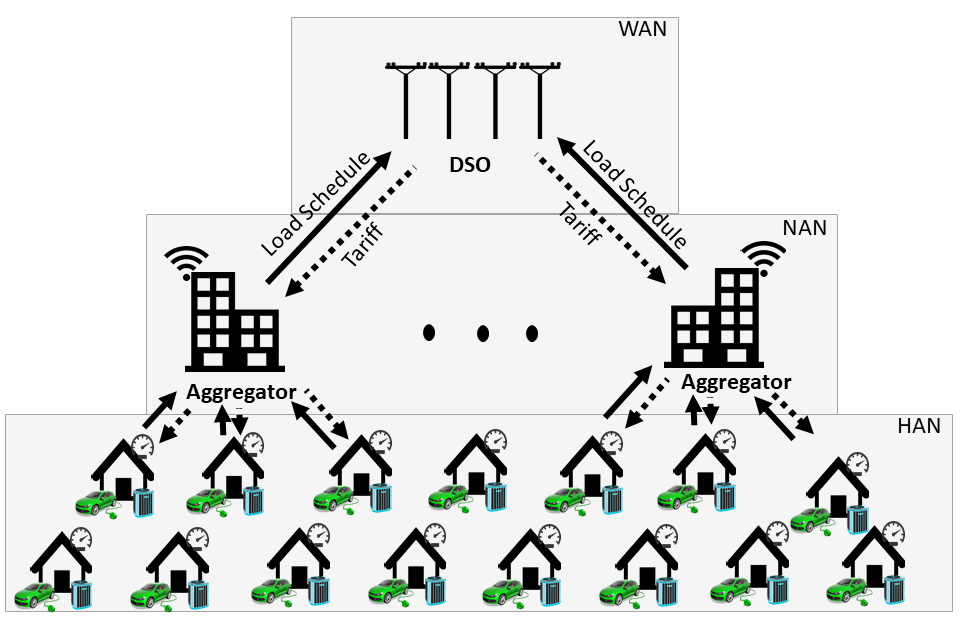}
  \caption{General representation of DSM-based congestion management}
  \label{fig:fig5}
\end{figure}

The general methodology adopted in market-based congestion management methods is illustrated in Figure \ref{fig:fig5}. Three entities are involved: DSO, aggregators, and consumers. DSO is the central system operator that is responsible for ensuring the absence of network congestion. It utilizes the Wide Area Network (WAN) to receive optimized load schedules from the aggregators, performs load flow analysis, and then determines the congestion tariff to be imposed to alleviate congestion. These tariffs are then sent to the aggregators to re-optimize their load schedules based on their customers' preferences. Aggregators are responsible for receiving their customers' preferences using the Neighborhood Area Network (NAN) and optimizing their load schedules to minimize their costs. Consumers are the owners of flexible and non-flexible loads and they hire aggregators to represent their needs in the electricity market. Utilizing the Home Area Network (HAN), consumers transmit their loads and their preferences to their respective aggregators.

DSM-based congestion management methods' reliance on the communication layer to relieve congestion makes them prone to cyber attacks. A Load Altering cyber Attack (LAA) can be implemented by compromising aggregators and altering the load schedules that are transmitted to the DSO. The optimal load schedule of aggregators can be defined as follows,
\setlength{\intextsep}{1pt}
\begin{equation}
\label{eqn1}
\begin{split}
P_{i,t}^{a}=E(P_{c,t}^{a,nf}+P_{c,t}^{a,f}),
\forall a \in N_a, \forall i \in N_d,\\ \forall c \in N_m, \forall t \in N_T \\
\end{split}
\end{equation}
where $P_{i,t}^{a}\in \mathbb{R}^{N_d}$ is the load of bus $i$ of aggregator $a$ at time $t$. $N_a$, $N_d$ and $N_m$ are the set of aggregators, buses, and customers in the network. $N_T$ is the total set of time slots considered in the day-ahead market. $E$ is the customer-to-bus mapping matrix where $E\in \mathbb{R}^{{N_d}\times{N_m}}$. $P_{c,t}^{a,nf}$ and $P_{c,t}^{a,f}$ are the non-flexible and flexible loads of customer, $c$, of aggregator $a$ at time $t$, where $\{P_{c,t}^{a,nf},P_{c,t}^{a,f}\}\in \mathbb{R}^{N_m}$. The load schedule of compromised aggregator can be defined as follows,
\setlength{\textfloatsep}{5pt}
\begin{equation}
\label{eqn1}
\begin{split}
P_{i,t}^{a^c}=P_{i,t}^{a^c}+\Delta_i^{a^c}\text{, for }0 \leq \Delta_i^{a^c} \leq \gamma
\end{split}
\end{equation}
where $a^c$ is the aggregator that is compromised and $\Delta_i^{a^c}$ represents the alteration of the compromised aggregator's load. To not raise any alarms with the DSO, $\Delta_i^{a^c}$ needs to be as low as possible. $\gamma$ is the upper limit for the change in the compromised aggregators' load schedule. In this paper, $\gamma$ is chosen as 0.1 to ensure the stealthiness of the attack while achieving the objective of the attacker. 

\begin{figure}[t]
\centering
  \includegraphics[trim={0cm 1cm 0cm 0cm}, width=0.3\textwidth]{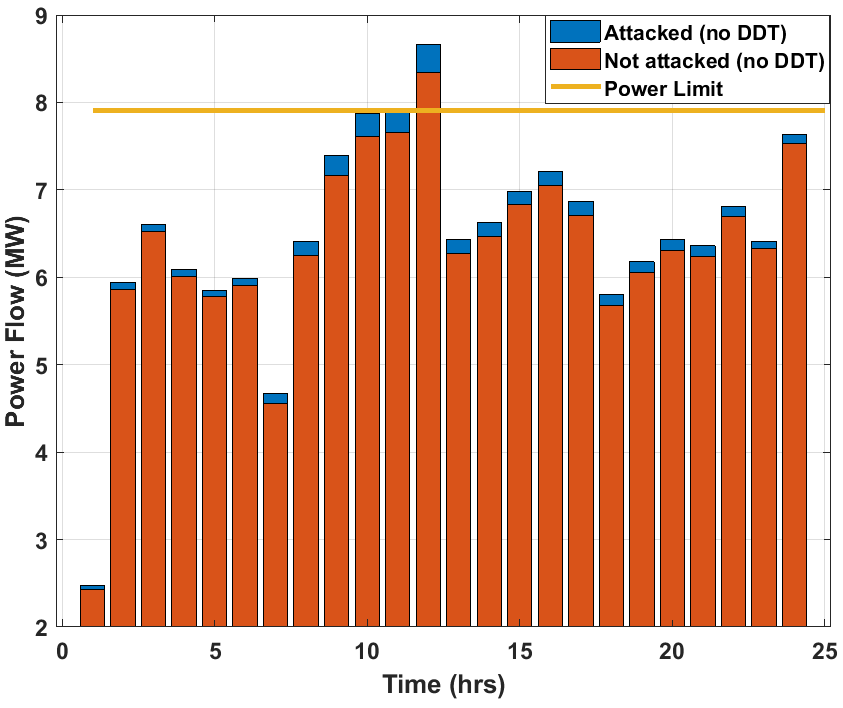}
  \caption{Effect of a LAA on power flow levels}
  \label{fig:fig51}
\end{figure}

\begin{figure}[t]
\centering
  \includegraphics[trim={0cm 1cm 0cm 0cm}, width=0.3\textwidth]{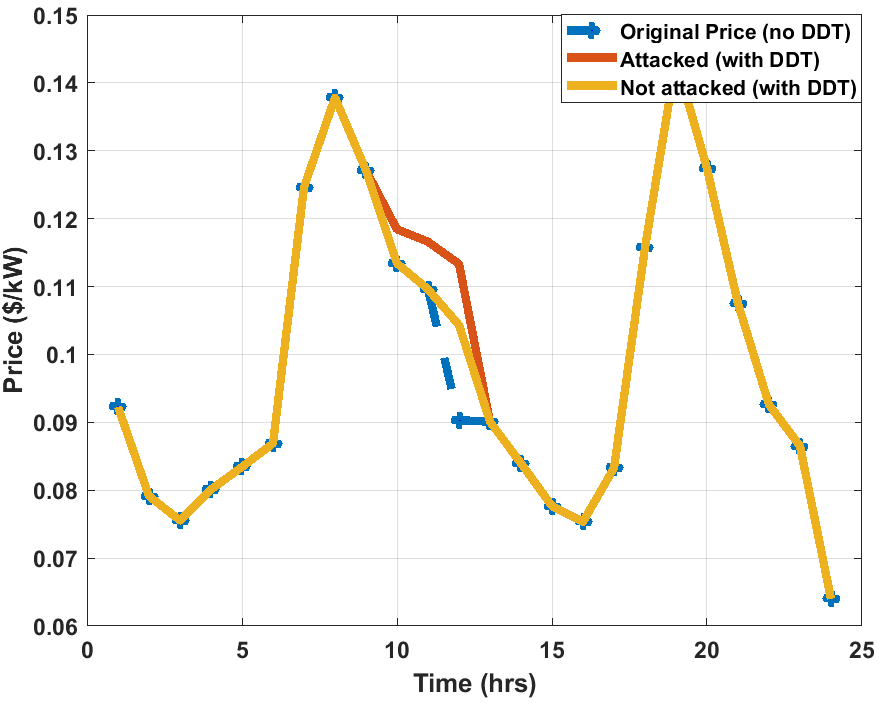}
  \caption{Effect of LAA on electricity price}
  \label{fig:fig52}
\end{figure}

Adopting the case study explained in Section II, and utilizing the DDT congestion management method \cite{b3} to mitigate congestion, the load schedules of aggregators 3 and 7 are compromised to study the impact of the cyber attack. Figure \ref{fig:fig51} illustrates the effect of LAA on power flow in line 0-1 of the IEEE 33 bus system as a comparison between the attacked and not attacked scenarios, prior to imposing any congestion tariff. At peak time, $t=12$, the line 0-1 was already congested. However, the LAA caused an increase in the congestion level. Moreover, line 0-1 was not congested at times 10 and 11. However, on scaling the load schedules of the two aggregators, congestion was created. This has an impact on electricity price as illustrated in Figure \ref{fig:fig52}. In the absence of a LAA, the imposed congestion tariff increased the price of electricity at t = 12 by 15\%. However, due to the load altering attack, the imposed tariff increased the price by a not needed 10\% at t = 12 which customers have to pay for a higher congestion level that is not real. Also, as a result of fake congestion being created at times 10 and 11, a congestion tariff is imposed by the DNO to solve the fake congestion. Customers now have to pay 4.4\% more at t = 10 and 6.4\% more at t = 11.

\section{Modeling of LAA Detection Scheme}
To detect LAAs, Recurrent Neural Networks (RNN), which is a class of Artificial Neural Networks (ANN), is proposed in this paper. Traditional backpropagation NNs, though capable of learning nonlinear functions, cannot capture sequential information in the input data which is required for dealing with sequence data. Neglecting data's temporal correlation impacts the model's accuracy \cite{jj}. RNNs, on the other hand, have a recurrent connection in their hidden states making them capable of capturing sequential information present in input data. The network, hence, develops a memory of past events and encodes it in its hidden state. RNNs are ideal for applications in which there is a temporal dependency in the data \cite{rnnbook}. Hence, an RNN's output at $t$ could have an effect on its decision at $t+1$. There are many types of RNNs based on their input and output data: sequence input (also known as many-to-one), sequence output (also known as one-to-many), sequence input and output (also known as many-to-many), or synced sequence inputs and outputs (also known as bidirectional many-to-many). RNNs are trained using backpropagation. However, for long sequences, the gradients, which carry information utilized for RNN parameter update, measured in the last layers becomes smaller. Thus, as a consequence parameter updates become insignificant. LSTM RNNs is an improvement over RNNs since it reduces RNNs vanishing gradient problem. LSTM's insensitivity to the gap between the relevant information in the past and the time when that information is needed in the present gives it an advantage over alternative sequential models, such as Hidden Markov Models \cite{rnnbook1}.

\begin{figure}[b]
\centering
  \includegraphics[trim={1cm 1cm 0cm 0cm}, width=0.45\textwidth]{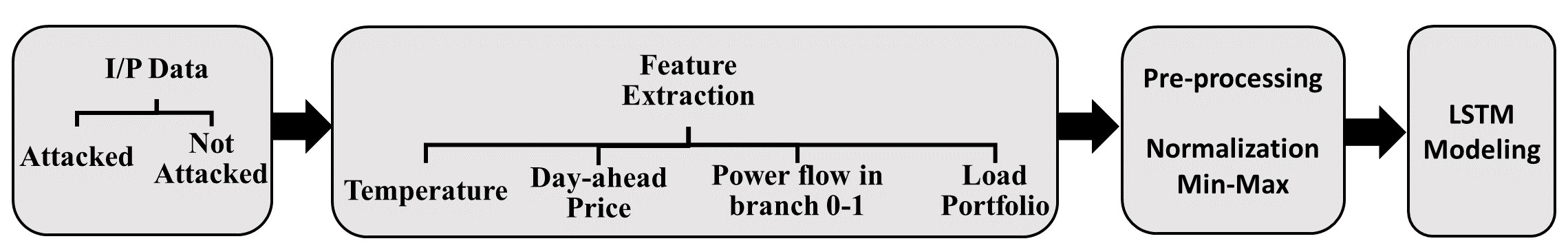}
  \caption{LSTM-based LAA Detection Model}
  \label{overall}
\end{figure}

Being in the day-ahead market, and receiving the aggregators' load schedules as a full day sequence makes utilization of LSTM as a cyber attack detection scheme attractive for the DSO prior to computation of the congestion tariff. To model the detection scheme, the steps illustrated in Figure \ref{overall}, and explained in the following subsections, needs to be executed.

\subsection{Data set, Feature Extraction and Preprocessing}
To model the LSTM network, hourly customer real power load, electricity price, and weather data for Michigan, USA was utilized. Data was collected for random days in November to March of 2015 to 2020. These winter months were chosen as the loads are at their peak, and there is a high chance of the network being congested. The real power load per bus in the distribution network is determined using the weather and day-ahead electricity prices to optimize aggregators' consumers' load. Hence, the four types of sequences that were chosen as input features are as follows:  
\begin{itemize}
    \item Historical daily weather data.
    \item Historical day-ahead electricity prices \cite{pjm}.
    \item Three days historical real power load per bus.
    \item Power load per bus.
    \item Power flow in the main feeder of the distribution network.
\end{itemize}
All the data were normalized using min-max normalization. For every feature, the minimum value was transformed to a 0, the maximum value was changed to a 1, and every other value gets represented by a decimal between 0 and 1. 1550 days of data were used to simulate attacked and not attacked scenarios. An attacked day is labeled as 1, and 0 otherwise.

\subsection{LSTM Modelling}
\begin{figure}[b]
\centering
  \includegraphics[trim={1cm 1cm 1cm 1cm}, width=0.35\textwidth]{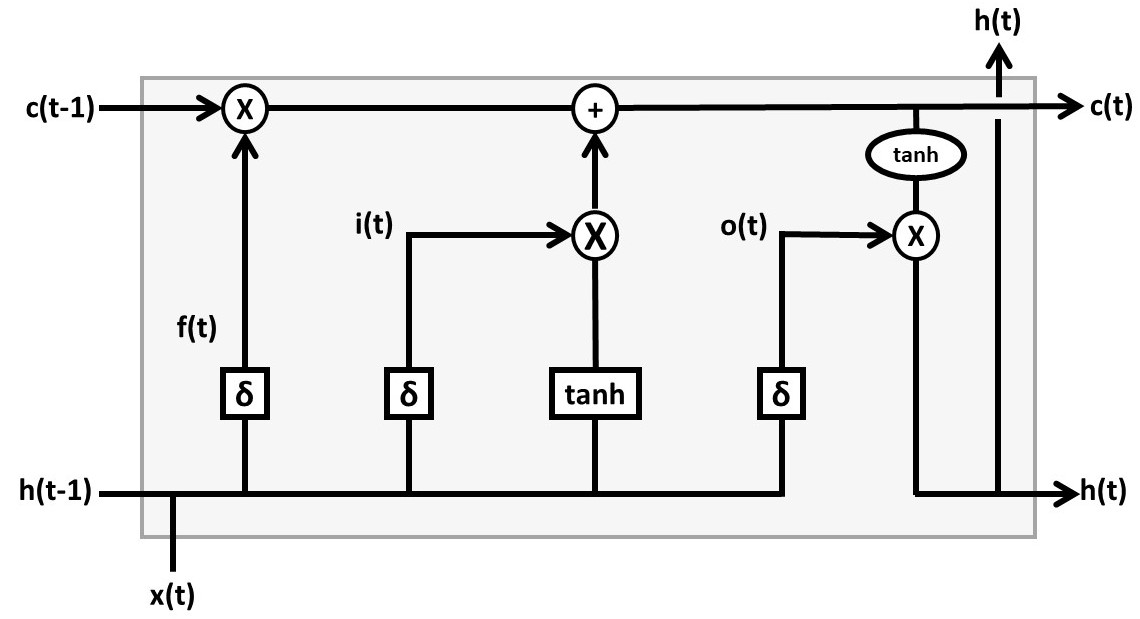}
  \caption{Structure of the LSTM-RNN memory-unit}
  \label{fig:fig53}
\end{figure}

To model LAA LSTM-based detection scheme, a many-to-one LSTM model is adopted. The input to the model is in the form of sequential data representing the 24-hour weather forecast, day-ahead electricity prices, real power load per bus, and the power flow in the feeder of concern. The output to the model, on the other hand, is a classification decision. A decision equal to $1$ represents an attacked scenario, while a decision equal to $0$ represents a not-attacked scenario. 
The LSTM-RNN model includes multiple LSTM-RNN memory units that replaces each node in the hidden layer of a neural network model. Each unit consists of an input gate, $i(t)$, an output gate, $o(t)$, a forget gate, $f(t)$, and a memory cell state, $c(t)$, as illustrated in Figure \ref{fig:fig53} \cite{lstm}. The three gates are responsible for selecting and rejecting the information that passes through the network in its cell state. Initially, the LSTM needs to determine which information it will remove from the cell state, $c(t-1)$. This is done by the sigmoid layer forget gate, $f(t)$, which takes its previous cell output, $h(t-1)$, and the new cell input, $x(t)$, to output a number between 0 and 1 for $c(t-1)$. A 1 indicates retaining the contents of $c(t-1)$ while a 0 means deleting the contents. 

\begin{figure}[b]
\centering
  \includegraphics[trim={1cm 1cm 1cm 1cm}, width=0.4\textwidth]{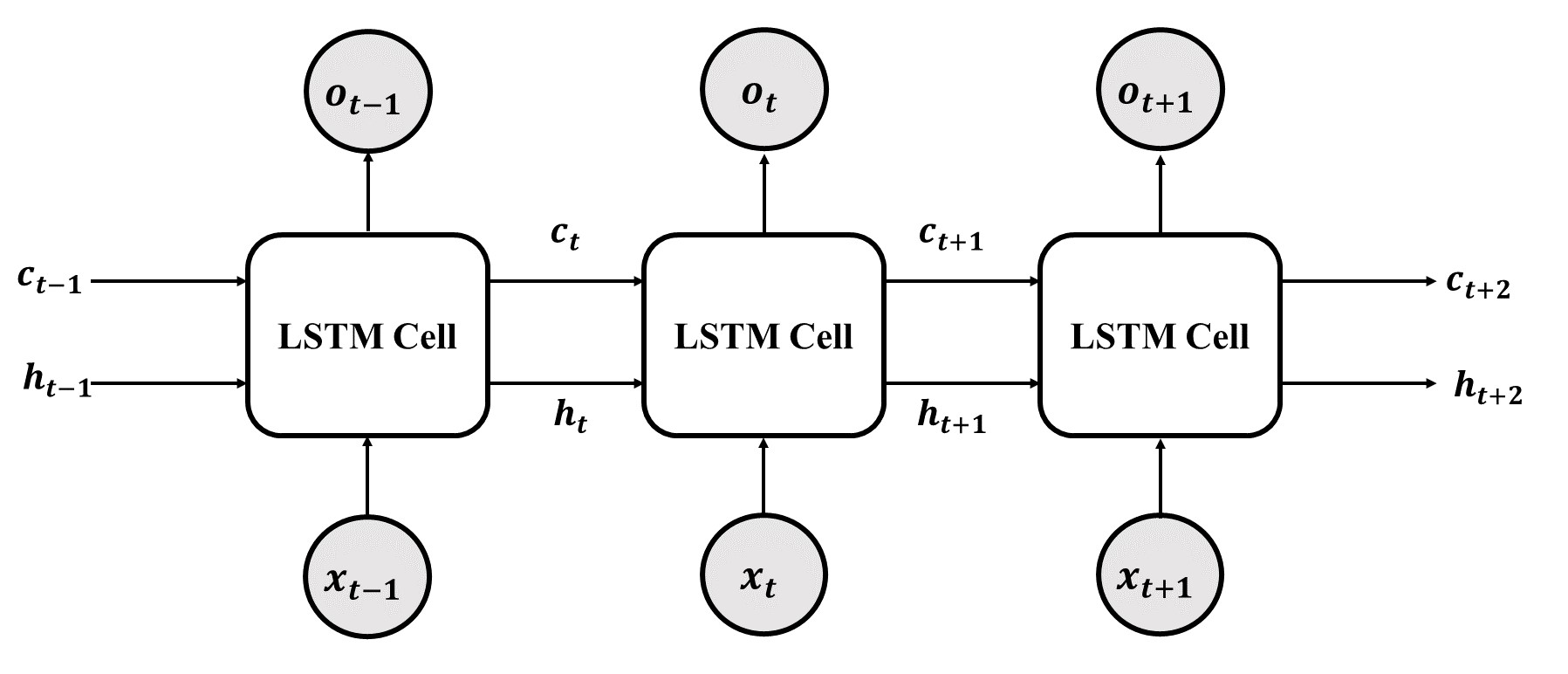}
  \caption{Overall Structure of the LSTM-RNN Model}
  \label{LSTMcycle}
\end{figure}

\begin{equation}
\label{eqn11}
f(t)=\delta(W_{f}[x(t),h(t-1)]+b_{f})\\ 
\end{equation}
A sigmoid layer input gate then takes the current input, $x(t)$, and previous cell output, $h(t-1)$, to determine which cell states to update. A tanh layer is responsible for creating an array of new updated values that would be added to the state. 
\begin{equation}
\label{eqn12}
i(t)=\delta(W_{i}[x(t),h(t-1)]+b_{i})\\ 
\end{equation}
\begin{equation}
\label{eqn13}
\Tilde{c}(t)=tanh(W_{c}[x(t)+h(t-1)+b_{c})\\
\end{equation}
The previous cell state, $c(t-1)$, is then updated by multiplying it with $f(t)$ (\ref{eqn11}) to forget the contents that was decided to be forgotten. The result is added to the product of $i(t)$ (\ref{eqn12}) and $\Tilde{c}(t)$ (\ref{eqn13}) to update the cell state values. 
\begin{equation}
\label{eqn14}
\Tilde{c}(t)=f(t)\times{c}(t-1)+i(t)\times\Tilde{c}(t)\\
\end{equation}
The output gate, $o(t)$, utilizes the sigmoid activation function to decide which part of the current cell makes it to the output. These values are then filtered using a tanh function to be between -1 and 1.
\begin{equation}
\label{eqn15}
o(t)=\delta(W_{o}[x(t),h(t-1)]+b_{o})\\ 
\end{equation}
\begin{equation}
\label{eqn16}
h_{t}=o_{t} \tanh{c(t)}\\
\end{equation}
$W_{f}$, $W_{i}$, $W_{c}$, and $W_o$ represent the weight matrices for the inputs of the network activation functions. Also, $b_{f}$, $b_{i}$, $b_{c}$, and $b_o$ represent the biases for the different gates. The above process is repeated for all time steps, as illustrated in Figure \ref{LSTMcycle}. Weights and biases are updated by minimizing the differences between the LSTM outputs and the training samples. 

\subsection{Results and Discussion}
$70\%$ of the data explained in the section earlier, corresponding to 1200 days of data representing attacked and not attacked scenarios, were used for training and validating the LAA detection scheme. The remaining $30\%$ of the data, corresponding to 350 days, were used for testing the performance of the model. the performance of the LSTM detection scheme was compared with the traditional back propagation neural network (ANN) model. To evaluate its performance, confusion matrix and accuracy were used as performance metrics. A confusion matrix assesses the performance of the model in terms of true positives, true negatives, false positives, and false negatives. False positives refer to the number of cases that were classified as attacked when in reality they were not attacked. On the other hand, a false negative refers to the number of cases that were not classified as attacked by the model. True positives and true negatives refer to case scenarios representing attacked and not attacked days respectively and the model classified it correctly. On the other hand, accuracy represents the correctness of the classifier and is calculated as follows:
    \begin{equation}
    \begin{split}
        \text{Accuracy }=\frac{\text{ (True Positive + True Negative)}}{\text{Number of Cases}}
    \end{split}
      \label{r}
    \end{equation}

Table \ref{table:table54} illustrates the confusion matrix of the results obtained using the LSTM and the ANN models. As observed, in both the models the number of false positive is higher compared to that of the false negative cases. The ANN resulted had nearly $9.4\%$ of false positive cases compared to $1.7\%$ of false negatives. Similarly, LSTM resulted in nearly $2.9\%$ of false positives compared to $0.6\%$ of false negatives. Using more data to train the model on more attack scenarios can decrease this number so that the model could clearly distinguish between an attacked and not attacked scenario. Compared to the ANN model, the LSTM model performed better. Summing up the number of true positive and true negative cases, as illustrated in (\ref{r}), results in an accuracy of $96.57\%$ for the LSTM model and accuracy of $88.86\%$ for the ANN model. The results signify the advantage of using LSTM sequential classifier to model the LAA detection scheme.

In the day-ahead market, on receipt of the aggregators' load schedules, the DSO executes the proposed LSTM model to detect if an attack has occurred. If the LSTM classifier detects no attack, then the DSO would go ahead to compute the congestion tariffs. However, if an attack is detected the DSO would inform the aggregators that a cyber attack has occurred in the network so that the aggregators could check their cyber security. Until the cyber attack is eliminated, the DSO would rely on its historical data to determine if there is congestion in the network. Future work should incorporate the detection of the aggregator that is compromised in the network.

 \setlength{\tabcolsep}{1pt}
 \begin{center}
\begin{table}[b]
\centering
\caption{Confusion Matrix of LAA Detection Models}
\begin{tabular}{|l||c|c||c|c|}
\hline
\multicolumn{1}{|c||}{} & \multicolumn{2}{c||}{ANN} & \multicolumn{2}{c|}{LSTM} \\
\cline{1-5}
\multicolumn{1}{|c||}{\diaghead{Output  Actual}{\textbf{Output  } }{\textbf{Actual}} } & \textbf{Attack} &\textbf{\makecell{No \\Attack}}& \textbf{Attack} &\textbf{\makecell{No \\Attack}}\\
\hline  \hline
\textbf{Attack}& 137 &33 & 141 & 10  \\ 
\textbf{No Attack} & 6 & 174 & 2 & 197\\ 
\hline\hline

\end{tabular}
\label{table:table54}
\end{table}
\end{center}

\section{Conclusions}
Cyber vulnerabilities of market-based congestion management techniques were studied in this paper. The impact of load altering attacks that creates fake congestion was evaluated in terms of congestion level and congestion tariffs. A detection technique using LSTM recurrent neural networks was proposed. On receiving aggregators' load schedules in the day-ahead market, the DSO would simulate the LSTM detection model to determine if any of the data received for each bus has been compromised. Distinctive sequences were chosen as features for the detection model to classify the presence or absence of an attack in the day-ahead market. Although the attacks simulated were made to be stealthy and hence hard to be detected, an accuracy of $96.57\%$ was obtained reflecting the capability of the proposed model in detecting load altering cyber attacks that compromises aggregators in the network. 

\bibliography{bibliography.bib}
\end{document}